\documentclass{article}

\usepackage{arxiv}

\usepackage[T1]{fontenc}    
\usepackage{hyperref}       
\usepackage{url}            
\usepackage{booktabs}       
\usepackage{amsfonts}       
\usepackage{nicefrac}       
\usepackage{microtype}      
\usepackage{lipsum}
\usepackage{graphicx}
\usepackage{float}
\usepackage{makecell}
\usepackage{csquotes}
\usepackage{subcaption}

\usepackage[utf8]{inputenc}
\usepackage{pgfplots}
\DeclareUnicodeCharacter{2212}{−}
\usepgfplotslibrary{groupplots,dateplot}
\usetikzlibrary{patterns,shapes.arrows}
\pgfplotsset{compat=newest}

\usepackage[
  style=alphabetic 
]{biblatex}
\date{February 3, 2022}

\title{Investigating Efficient Deep Learning Architectures For Side-Channel Attacks on AES}

\author{
Yohaï-Eliel BERREBY \\
  Télécom Paris\\
  \texttt{me@yberreby.com}\\
  \And
Laurent SAUVAGE \\
  Télécom Paris\\
  \texttt{laurent.sauvage@telecom-paris.fr}
}

\begin{document}
\maketitle
\begin{abstract}
  Over the past few years, deep learning has been getting progressively more
  popular for the exploitation of side-channel vulnerabilities in embedded
  cryptographic applications, as it offers advantages in terms of the amount of
  attack traces required for effective key recovery. A number of effective
  attacks using neural networks have already been published, but reducing their
  cost in terms of the amount of computing resources and data required is an ever-present
  goal, which we pursue in this work.

  This project focuses on the ANSSI Side-Channel Attack Database (ASCAD).  We
  produce a JAX-based framework for deep-learning-based SCA, with which we
  reproduce a selection of previous results and build upon them in an attempt
  to improve their performance. We also investigate the effectiveness of
  various Transformer-based models.
\end{abstract}

\keywords{
  Deep Learning \and Cryptography \and Side-Channel Attacks \and
  Profiling Attacks
}

\section{Introduction}

\subsection{Context: Side-Channel Attacks}

Side-Channel Attacks (or SCA) are a class of cyberattacks that exploit
weaknesses specific to implementions of would-be secure systems. For information
recovery, they may rely on correlation between the targeted data and variations
in timing \cite{kocher1996timing}, power consumption and electromagnetic (EM)
emissions \cite{randolph2020power}, sound emission \cite{genkin2017acoustic},
and other characteristics of a system. They may rely on the system's failure
behavior, through Differential Fault Analysis \cite{dusart2003differential}. If
a suitable side-channel vulnerability exists in its implementation(s), even a
fully theoretically-secure algorithm can be cracked.

The implementer of a system may try to minimize information leakage through
\emph{countermeasures}. These may include performing sensitive operations in
constant time; always executing the same code regardless of the input fed to
the system; avoiding the direct manipulation of sensitive data through
\emph{masking}; etc. They may also include physical defense mechanisms, such as
EM shielding.

Side-channel attacks may be \emph{profiling} or \emph{non-profiling}, the
difference being that the attacker has access to a copy of the target device in
the \emph{profiling} case. In this work, we will focus on \emph{profiling}
attacks, into which machine learning techniques have been making headway. In
particular, neural networks have been studied due to their ability to recover
information even from highly-protected implementations.

As new attacks are published, increasingly-effective countermeasures are
developed, rendering further attacks more costly both in terms of computing
power required to train the models, and in terms of the amount of data that must
be collected in order to apply them. As such, enhancing the information-recovery
capabilities of neural networks used for SCA is of major interest. 

So far, in practice, even neural networks are typically unable to reliably
recover the correct value of a given key byte from a single trace. As such,
guesses made over multiple traces are combined to obtain a more reliable
one. The minimum number of traces required for reliable recovery of the target
variable is commonly referred to as "guessing entropy" in the literature.

In this project, we focus on power consumption and EM traces. We use the ASCAD
project, described below, as a source of data and as a starting point.

\subsection{Overview of ASCAD}

ASCAD (ANSSI SCA Database) \cite{ascadgit} is a collection of power trace
databases for side-channel attacks. Since the introduction of its first version,
has enjoyed significant popularity in the SCA community as a benchmark for
deep-learning-based attacks.

Thus far, it comprises two main versions,
both targeting AES and collected on different microcontrollers
and AES implementations. The project comes with a set of pre-trained models
targeting each database, as well as code to train them. 

Each set of databases is provided with "raw" traces, covering some portion of
the encryption/decryption process, and "extracted" traces, covering a subset of
the former which is known to leak relevant information. Extracted
traces come with precomputed \emph{labels}, corresponding to intermediate
variables manipulated during the encryption/decryption process from which the
key can be recovered. It is significantly easier to recover a well-chosen
intermediate variable then post-process it to compute the key than to try to
recover the key directly.

Traces may be synchronized or desynchronized. In the synchronized case, a given
timestep always corresponds to the same instant in the encryption or decryption
process, whereas such an instant may be represented at different time steps in
the desynchronized case. Desynchronization implies the need for some degree of
shift invariance or equivariance in the network's feature extraction process.

\subsubsection{ASCADv1 - Implementation on ATMega8515}

Described in \cite{ascadv1}, ASCADv1 has two campaigns (themselves ambiguously
named v1 and v2 - we refer to them as "fixed key" and "variable
key") targeting a software AES implementation on ATMega8515 \cite{secAESATM},
which uses boolean masking.

The fixed-key campaign uses the same key for profiling and attack sets, whereas
the variable-key campaign uses a random key for profiling and a fixed one for
attack. 

They are structured as shown in \autoref{tab:ascadv1layout}.

\begin{table}[H]
  \caption{ASCADv1 (ATMega) datasets (SPET/SPRT = Samples Per Extracted/Raw Trace)}
  \centering
  \begin{tabular}{|l|l|l|l|l|l|}
  \hline
  \thead{Version}
    & \thead{N° Profiling traces}
    & \thead{N° Attack traces}
    & \thead{SPET}
    & \thead{SPRT} \\
  \hline
  v1 / fixed key 
    & 50,000
    & 10,000
    & 700 
    & 100,000
 \\
  \hline
  v2 / variable key
    & 200,000
    & 100,000
    & 1400
    & 250,000
    \\
  \hline
  \end{tabular}

  \label{tab:ascadv1layout}
\end{table}

The README file for the variable-key campaign mentions that the traces are "not
synchronized". The traces do appear synchronized \emph{to some degree}; the
magnitude of the desynchronization is not explicited.

\begin{figure}
  \center
\begin{tikzpicture}

\definecolor{darkgray176}{RGB}{176,176,176}
\definecolor{lightgray204}{RGB}{204,204,204}

\begin{axis}[
legend style={fill opacity=0.8, draw opacity=1, text opacity=1, draw=lightgray204},
tick align=outside,
tick pos=left,
x grid style={darkgray176},
xlabel={Time},
xmin=-34.95, xmax=733.95,
xtick style={color=black},
y grid style={darkgray176},
ymin=-68.05, ymax=43.05,
ytick style={color=black}
]
\addplot [semithick, red, forget plot]
table {%
0 20
1 19
2 13
3 6
4 -2
5 -10
6 -19
7 -29
8 -39
9 -47
10 -52
11 -53
12 -52
13 -49
14 -45
15 -42
16 -38
17 -33
18 -24
19 -13
20 -2
21 7
22 13
23 15
24 15
25 15
26 15
27 14
28 13
29 10
30 4
31 -4
32 -12
33 -20
34 -27
35 -33
36 -39
37 -45
38 -48
39 -46
40 -39
41 -24
42 -10
43 3
44 10
45 11
46 8
47 4
48 3
49 6
50 11
51 16
52 18
53 17
54 12
55 5
56 -3
57 -11
58 -21
59 -31
60 -40
61 -48
62 -51
63 -52
64 -50
65 -46
66 -42
67 -40
68 -38
69 -33
70 -26
71 -15
72 -5
73 4
74 10
75 12
76 12
77 12
78 11
79 11
80 9
81 6
82 -1
83 -7
84 -15
85 -22
86 -28
87 -33
88 -38
89 -44
90 -47
91 -46
92 -40
93 -29
94 -16
95 -4
96 2
97 4
98 1
99 -2
100 -3
101 1
102 5
103 10
104 11
105 9
106 5
107 0
108 -7
109 -14
110 -21
111 -30
112 -38
113 -46
114 -50
115 -51
116 -49
117 -46
118 -43
119 -41
120 -39
121 -33
122 -26
123 -15
124 -6
125 3
126 7
127 9
128 9
129 7
130 7
131 7
132 6
133 4
134 -2
135 -9
136 -17
137 -23
138 -29
139 -34
140 -39
141 -43
142 -46
143 -43
144 -32
145 -16
146 5
147 24
148 35
149 38
150 33
151 28
152 22
153 22
154 24
155 27
156 29
157 26
158 20
159 10
160 0
161 -12
162 -24
163 -37
164 -49
165 -59
166 -63
167 -63
168 -58
169 -51
170 -44
171 -39
172 -34
173 -29
174 -21
175 -11
176 1
177 12
178 20
179 25
180 27
181 26
182 25
183 23
184 19
185 13
186 4
187 -5
188 -15
189 -24
190 -31
191 -38
192 -43
193 -48
194 -53
195 -54
196 -47
197 -35
198 -19
199 -2
200 11
201 18
202 18
203 16
204 15
205 16
206 20
207 24
208 27
209 26
210 21
211 14
212 5
213 -5
214 -15
215 -27
216 -37
217 -48
218 -56
219 -61
220 -60
221 -56
222 -52
223 -48
224 -43
225 -38
226 -31
227 -21
228 -10
229 0
230 8
231 13
232 15
233 16
234 17
235 17
236 17
237 14
238 9
239 0
240 -9
241 -18
242 -25
243 -32
244 -38
245 -44
246 -48
247 -48
248 -43
249 -33
250 -19
251 -6
252 2
253 6
254 4
255 2
256 1
257 3
258 8
259 14
260 17
261 17
262 13
263 7
264 0
265 -8
266 -16
267 -25
268 -35
269 -43
270 -49
271 -51
272 -49
273 -47
274 -45
275 -43
276 -40
277 -36
278 -29
279 -20
280 -10
281 0
282 6
283 8
284 9
285 8
286 9
287 9
288 9
289 7
290 1
291 -5
292 -12
293 -19
294 -26
295 -31
296 -37
297 -42
298 -45
299 -44
300 -36
301 -21
302 -3
303 13
304 23
305 27
306 25
307 19
308 15
309 14
310 16
311 20
312 23
313 22
314 17
315 10
316 0
317 -11
318 -21
319 -33
320 -43
321 -52
322 -58
323 -58
324 -55
325 -50
326 -44
327 -40
328 -36
329 -31
330 -25
331 -14
332 -3
333 7
334 15
335 19
336 20
337 20
338 19
339 18
340 15
341 11
342 4
343 -4
344 -14
345 -23
346 -29
347 -35
348 -40
349 -45
350 -49
351 -50
352 -47
353 -38
354 -25
355 -13
356 -4
357 0
358 -1
359 -2
360 -2
361 1
362 5
363 11
364 13
365 13
366 9
367 4
368 -2
369 -9
370 -16
371 -25
372 -33
373 -42
374 -48
375 -51
376 -51
377 -48
378 -46
379 -44
380 -42
381 -37
382 -30
383 -20
384 -9
385 0
386 7
387 9
388 8
389 8
390 8
391 9
392 9
393 7
394 3
395 -3
396 -11
397 -18
398 -25
399 -31
400 -37
401 -42
402 -46
403 -46
404 -38
405 -26
406 -9
407 7
408 19
409 24
410 20
411 15
412 11
413 11
414 14
415 18
416 21
417 21
418 17
419 10
420 0
421 -9
422 -20
423 -31
424 -41
425 -51
426 -56
427 -57
428 -55
429 -50
430 -45
431 -41
432 -37
433 -34
434 -26
435 -17
436 -6
437 4
438 12
439 17
440 19
441 19
442 17
443 16
444 14
445 10
446 4
447 -4
448 -13
449 -21
450 -28
451 -34
452 -39
453 -44
454 -49
455 -51
456 -46
457 -38
458 -25
459 -12
460 -2
461 2
462 2
463 0
464 -1
465 1
466 5
467 11
468 14
469 14
470 11
471 5
472 -2
473 -9
474 -16
475 -25
476 -35
477 -42
478 -50
479 -52
480 -52
481 -50
482 -47
483 -45
484 -42
485 -38
486 -30
487 -20
488 -9
489 1
490 8
491 11
492 11
493 10
494 10
495 11
496 11
497 9
498 4
499 -2
500 -10
501 -17
502 -25
503 -31
504 -36
505 -42
506 -46
507 -47
508 -41
509 -30
510 -14
511 3
512 16
513 23
514 23
515 19
516 15
517 14
518 16
519 19
520 22
521 23
522 19
523 13
524 4
525 -6
526 -16
527 -28
528 -39
529 -49
530 -56
531 -58
532 -57
533 -53
534 -49
535 -43
536 -40
537 -35
538 -28
539 -19
540 -9
541 2
542 10
543 15
544 17
545 17
546 18
547 18
548 15
549 12
550 5
551 -2
552 -11
553 -20
554 -28
555 -34
556 -39
557 -44
558 -49
559 -50
560 -46
561 -37
562 -23
563 -11
564 0
565 5
566 5
567 3
568 2
569 3
570 7
571 12
572 16
573 16
574 13
575 7
576 -1
577 -8
578 -16
579 -26
580 -35
581 -44
582 -51
583 -54
584 -53
585 -51
586 -48
587 -46
588 -43
589 -39
590 -33
591 -22
592 -11
593 0
594 7
595 11
596 12
597 11
598 11
599 10
600 10
601 9
602 5
603 -2
604 -10
605 -18
606 -24
607 -31
608 -37
609 -43
610 -48
611 -49
612 -45
613 -33
614 -17
615 1
616 13
617 20
618 19
619 14
620 10
621 10
622 12
623 17
624 21
625 21
626 18
627 12
628 3
629 -6
630 -17
631 -27
632 -38
633 -48
634 -56
635 -59
636 -57
637 -53
638 -48
639 -44
640 -40
641 -37
642 -31
643 -22
644 -11
645 0
646 9
647 15
648 17
649 17
650 16
651 15
652 14
653 11
654 6
655 -2
656 -11
657 -19
658 -27
659 -33
660 -38
661 -44
662 -48
663 -50
664 -46
665 -35
666 -19
667 0
668 14
669 22
670 22
671 18
672 15
673 14
674 16
675 21
676 25
677 25
678 21
679 15
680 5
681 -4
682 -15
683 -25
684 -37
685 -48
686 -56
687 -60
688 -59
689 -55
690 -50
691 -45
692 -41
693 -37
694 -32
695 -22
696 -11
697 0
698 9
699 15
};
\end{axis}

\end{tikzpicture}

  \input{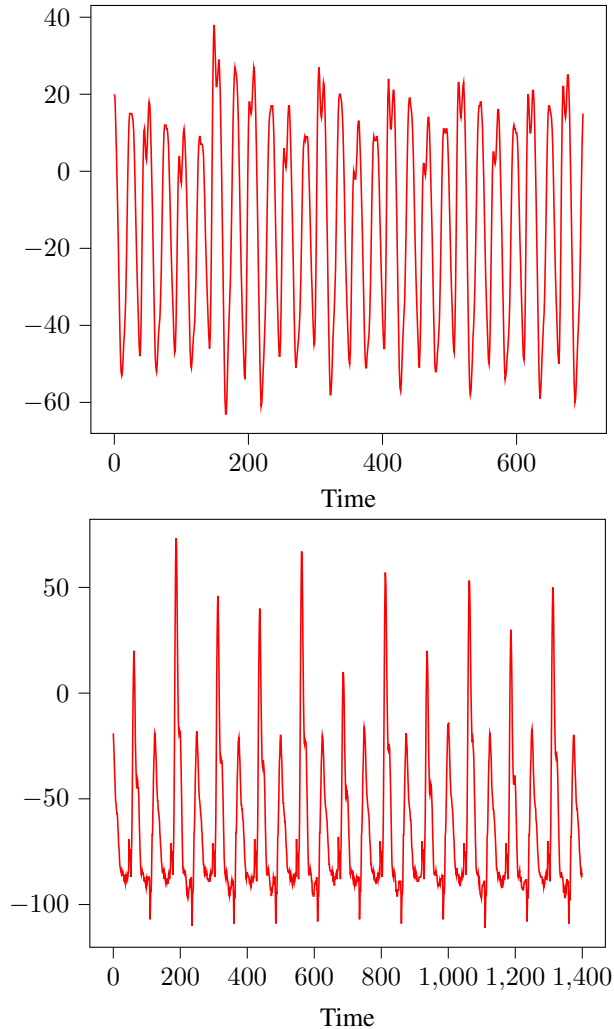}

  \caption{Comparison of traces between ASCAD fixed-key and variable-key ATMega
  datasets}

  \label{fig:fixed-var-comp}
\end{figure}

Additionally, the variable key dataset exhibits significant qualitative
differences from the fixed key dataset, as one can see on
\autoref{fig:fixed-var-comp}.

These differences were not clearly documented: no mention of the variable-key
dataset was made in \cite{ascadv1}, and no explanation was provided in the
repository.

They were noticed by other researchers, but clarification thus far
has been incomplete \cite{ascadIssue13} regarding the precise method of
measurement for each campaign.
Nevertheless, the difference in goals behind each campaign is to be noted:

\begin{displayquote}
 "The fixed key campaign was measured with a strong incentive to get a clean
 signal in order to make sure that simple attacks could be performed, while this
 effort was not stressed in the random key campaign, which aims at being more
 challenging. This explains the notable difference of sampling rate and signal
 quality that you observed."

 - @rb-anssi (Ryad Benadjila)
\end{displayquote}

The lack of clarity around the nature of the datasets, along with the relative
scarcity of papers exploiting the variable-key one, represented a significant
source of confusion at the beginning of this project.
It was originally incorrectly assumed that published attacks focused on the
variable-key dataset, as the performance of a network on the fixed-key
dataset said little about its ability to generalize to different keys.

\subsubsection{ASCADv2 - Implementation on STM32F303RCT7}

Described in \cite{ascadv2}, ASCADv2 has a single campaign, with a total of
800,000 traces. The "extracted" dataset features a profiling set of 500,000
traces using random keys, and an attack set of 10,000 traces using random keys
as well. The paper claims 700,000 profiling traces, 20,000 validation traces and
50,000 test traces, but the data linked on the project's GitHub appears to be
different from the one it describes.

The AES implementation it targets \cite{secAESSTM} uses affine masking
\cite{affineMasking} as a side-channel countermeasure instead of the simpler
boolean masking technique used on the ATMega8515.

Except for data exploration, our tests were not applied to ASCADv2 over the
course of this project. We had initially hoped to do so, but ASCADv1 by itself
presented difficulties that we wanted to resolve first.

\subsection{Prior work on ASCAD}

Ever since ASCAD's introduction, there has been a notable body of research on
efficient architectures for attacks on the database, in addition to the ones put
forth in the original paper. We chose to focus on a handful of them, described
below.

\subsubsection{Original ASCADv1 networks}

\cite{ascadv1} explored Multi-Layer Perceptrons (MLPs), and several Convolutional Neural Network (CNN) architectures: VGG-16 \cite{vgg},
ResNet-50 \cite{resnet} and Inception-v3 \cite{inceptionv3}.

CNNs and MLPs had comparable performance in the synchronized key, but CNNs
vastly outperformed MLPs in the presence of the desynchronization, owing to
the translation equivariance of convolutions. Among CNNs, a VGG-16-inspired
network had the best accuracy.

Training setup:

\begin{itemize}
  \item Optimizer: RMSProp
  \item Learning rate schedule: constant at $10^{-5}$
  \item Preprocessing: none (unscaled, uncentered traces)
\end{itemize}

\subsubsection{"\textit{Methodology for Efficient CNN Architectures in Profiling Attacks}"}

\cite{cnnMethodology} put forth a disciplined methodology to build efficient
CNNs for profiling attacks. It uses relatively shallow architectures, with short
filters, leverages batch normalization \cite{bn} and a one-cycle learning rate
schedule.

In parallel, the authors used gradient \cite{gradVisMasure} and activation
visualization to pinpoint the timesteps considered to be of interest by the
network, and compared them to Signal-to-Noise Ratio (SNR)
analyses on intermediate variables of interest.

For ASCADv1, its best network in the synchronized case has 3,930 times fewer
parameters than the previous state of the art, and has a greater information recovery
capability, requiring 191 traces for a zero-entropy key recovery, against a
previous best of 1,146. Only the fixed-key dataset was investigated by the
authors.

Training setup:

\begin{itemize}
  \item Optimizer: Adam \cite{adam}
  \item Learning rate schedule: linear one-cycle with maximum of $10^{-3}$ 
  \item Preprocessing: point-wise (synchronized) / none (desynchronized)
\end{itemize}

\subsubsection{"\textit{Pay Attention to Raw Traces: A Deep Learning Architecture for End-to-End Profiling Attacks}"}

The approach proposed in \cite{PART} (P.A.R.T.) sets itself apart from others by
being able to take "raw" traces as input - that is, traces covering a
significant part of the encryption/decryption cycle, and not just a small window
in which there is high information leakage. In that respect, it does away with
the need for the selection of Points Of Interest (POIs) prior to network
training and evaluation.

In order to be able to achieve that goal without running into the so-called
"curse of dimensionality" and the accompanying explosion in resource
consumption, P.A.R.T. first employs a so-called "junior encoder", which encodes
overlapping time windows with a width of 1-2 clock cycles into a much shorter
sequence:

\paragraph{Synchronized setting} Two sequences of time windows are encoded by
locally-connected layers with stride equal to their width, a single filter and
an offset of half their width between them. They are then concatenated. This
significantly reduces the dimensionality of the output fed to subsequent layers.

\paragraph{Desynchronized setting} A stride of 1 is used for the first layer, which
leads to no dimensionality reduction besides that ensuing from the lack of
padding. This reduction is achieved by stacking convolutional layers with
kernels of length 3 and a stride of 1 still, and max-pooling layers.
The final number of channels is 128.
Data augmentation is applied by randomly shifting the input in addition to its
original desynchronization.

After the junior encoder, a "senior encoder" is used to combine features across
time, using bidirectional LSTMs. It is followed by a simple multiplicative
attention mechanism and a feed-forward classifier.

Once fully converged, the authors' networks can recover a key from both datasets
in under 10 traces.

N.B.: In this paper, the ASCADv1 (ATMega) fixed-key dataset is described as "ASCAD v1",
and the variable-key dataset as "ASCAD v2".

\subsubsection{Original ASCADv2 network}

The neural network proposed in \cite{ascadv2} is a derivative of ResNet
\cite{resnet}. It uses multi-task learning to recover information about the
masking parameters and about each key byte simultaneously, with a common trunk
for feature extraction followed by branches specific to each predicted variable.

\subsection{Transformer and derivatives}

The introduction of the Transformer architecture in \cite{AIAYN}, whose
fundamental characteristic is the pervasive use of attention, represented a
major leap forward in deep learning. Since then, its derivatives have been
successfully applied to Natural Language Processing (NLP)
\cite{bert,gpt2,gpt3}, Computer Vision \cite{vit,cvt}, speech recognition
\cite{audiomer,gongAST}, symbolic computation \cite{sym}, etc., regularly
producing new state-of-the-art results. The vast flexibility of this family of
architectures made it a natural candidature for evaluation in the SCA context,
which, to our knowledge, had not been done before.

Transformer models take a (potentially variable-length) sequence of
\emph{tokens} as input, and return a sequence with the same shape.
To perform classification, this sequence may then be aggregated into a single
token, for example through global average pooling, or through extraction of the
embedding corresponding to a special, learned token, such as BERT's
\texttt{[CLS]} \cite{bert}.

Below, we detail some of the considerations specific to Transformers.

\subsection{Positional encoding}
By itself, the attention mechanism cannot distinguish between positions in the
input sequence. As such, additional features, called \emph{positional encoding},
are typically added or concatenated to the input embeddings before they are fed
to a Transformer trunk. They may be learnable, or deterministic, for example
using Fourier features as in the original Transformer paper. 

A significant number of such features may be required for the model to
accurately leverage positional information; it is typically on the order of a
few hundreds. This represents a significant overhead if the input tokens have
low dimensionality.

Some models, such as CvT \cite{cvt} or Audiomer \cite{audiomer}, do away with
positional encodings altogether by leveraging the spatially-aware inductive bias
of convolutions, applying them to tokens repeatedly within the Transformer
trunk.

\subsubsection{Training setup}

A major drawback of Transformer models thus far has been their high cost of
training. Though they do not necessarily have more trainable parameters than
comparable alternatives, the most popular Transformers do, and were designed
with very-large-scale training setups in mind. Derivative works rely on
fine-tuning more often than not, due to the unfeasibility of training those
models from scratch on a modestly-sized infrastructure.

One reason for this high cost is the amount of data required by Transformers to
generalize well, which is typically extremely large: BERT \cite{bert} was
pretrained on 3.3 billion words, and GPT-3 \cite{gpt3} on nearly 500 billion
tokens; ViT \cite{vit} performs much worse than ResNet when pretrained on a
dataset of 9 million images, but better when over 90 million images are used.
This may be explained by the flexibility of the attention mechanism, which is
both a strength and a weakness, as it lacks the inductive bias that led CNNs to
revolutionize computer vision in the early 2010s.

To train Transformers in the face of a small dataset, data augmentation is
nearly a requirement. While various modalities of data augmentation have been
extensively studied and implemented in the vision and audio subfields of machine
learning, they may be delicate to implement in the SCA context - beyond simple
desynchronization - without destroying the relevant information carried by
traces.

Additionally, Transformers are trained with adaptive optimizers such as Adam
\cite{adam}, LARS \cite{lars} or LAMB \cite{LAMB}.
Use of a learning rate schedule including warm-up is a necessity to ensure
stability, and proper hyperparameter tuning is recommended. Detailed training
guidelines may be found in \cite{trtr}.

\subsubsection{Taming quadratic complexity}

Another reason for Transformers' high training cost is the $O(n^2)$ time and
space complexity of the self-attention mechanism with respect to the length of
the input sequence, which typically renders it impractical for sequences longer
than a few hundred or a few (low single digits) thousand entries.

A variety of methods exist to work around the latter problem; we describe them below.

\paragraph{Artificial reduction of the sequence length}
This is the standard, most common workaround. For signal processing,
this may be done by splitting the input into potentially-overlapping patches of a
set size \cite{vit}, often with additional preprocessing such as Short-Time Fourier
Transforms (STFTs) \cite{gongAST,tinytrans}.

\paragraph{Performing self-attention on a fixed-size latent array}
Perceiver \cite{perceiver} and Perceiver IO \cite{perceiverIO} by DeepMind use
a novel architecture, wherein the model's trunk transforms a latent array whose
initial value is learned, and whose size is independent of the input sequence.
Cross-attention is performed between the input and the latent array several
times across a forward pass, but less often than self-attention on the latent array.
The resulting model has $O(m(m+n))$ complexity, where $m$ is the length of the
latent array and $n$ the length of the input. This architecture is particularly
relevant for multi-modality processing (video, audio, sound) with a single
model, as modalities can be distinguished using their additional dimensions in
the embeddings of the corresponding data.

\paragraph{Reducing the complexity of the attention mechanism itself}
This approach has notably been proposed in the Reformer \cite{reformer} and
Performer \cite{performer} papers. Reformers use a memory-efficient training
process by computing attention scores separately for each query, to avoid
storing a full attention matrix, and leverage locality-sensitive hashing to
only compute attention scores against queries that are relatively close a given
value; they achieve quasi-linear ($O(n \log n)$) complexity. Performers
approximate kernelizable attention mechanisms using a method called FAVOR+
(\emph{Fast Attention Via Positive Orthogonal Random Features}) and achieve
$O(n)$ complexity. In addition to its linear complexity, FAVOR+ has the
advantage of being a drop-in replacement to regular attention, with no
adjustments required besides switching out the attention function. Let it be
noted that the Audiomer models \cite{audiomer} use FAVOR+ in order to be
computationally tractable.

\subsection{Discussion and project direction}

As mentioned before, we decided to evaluate the effectiveness of Transformer
models in the SCA context. After the initial literature review, the project's
initial aim was to build upon P.A.R.T. \cite{PART}, replacing the senior
encoder's LSTMs with Transformer blocks using FAVOR+ attentions to achieve
linear complexity, and adapting the architecture's tenets to ASCADv2.

We decided to proceed using JAX \cite{jax2018github}, which is being positioned by
Alphabet as a long-term replacement for Tensorflow, offers significant
flexibility due to its ability to perform automatic differentiation (autodiff)
on Numpy-like Python code, and delivers high performance thanks to Just-In-Time
(JIT) compilation leveraging the XLA compiler.
Since JAX, by itself, doesn't offer common neural network primitives such as
batch normalization, convolutional layers, trainable parameter representation,
etc., we used DeepMind's Haiku library \cite{haiku2020github} on top of it.
Haiku is a thin abstraction layer, and is completed by composing parts of the
accompanying ecosystem; for example, optimizers are not included in Haiku, and
may be found in Optax \cite{optax2020github}. This compositional,
build-as-you-go setup involved significantly more friction than using
Keras/Tensorflow, and the quickly-evolving, young nature of the JAX ecosystem
meant that documentation was often scarce or outdated. In counterpart, this
proved to be an excellent learning opportunity, and it was possible to directly
leverage Google/DeepMind's latest research projects.

Regarding the training environment, some of the projects we intended to build
upon incurred significant training overhead (we measured 30 min / epoch on a
single NVIDIA V100 GPU for an ASCADv2 CNN, with training being prolonged to 300
epochs in the corresponding paper, and Transformers are known to be costly to
train and memory-hungry). As such, parallel training on powerful GPUs was a must.
To this end, our team secured a resource allocation of 10,000 GPU hours on the
Jean Zay supercomputer, part of the CNRS's IDRIS scientific computing platform.
This allowed us to train on up to eight 32GB V100 GPUs per node simultaneously, with
an upper limit of forty GPUs used simultaneously. Access was only fully acquired in
early December; before then, training was conducted on a single NVIDIA GTX 1660
Ti.

\section{Our contribution}

A significant portion of the time allocated to this project was dedicated to
becoming familiar with its foundations: Side-Channel Attacks, the ASCAD project
and surrounding literature, the JAX/Haiku ecosystem, Transformer models and
their variations, the Jean Zay scientific computing platform, parallel
training, regularization techniques, etc.

In addition to this foundational work, we ran experiments using the following
architectures (all reimplementations and adaptations were performed over the
course of this project):

\begin{itemize}
  \item Reimplementation of \cite{ascadv1}'s best synchronized CNN architecture
    (VGG-16-based) in JAX from Tensorflow
  \item Reimplementation of \cite{cnnMethodology}'s best synchronized
    architecture in JAX from Tensorflow, and experiments with variations of it
  \item Adaptation of Perceiver \cite{perceiver} / Perceiver IO
    \cite{perceiverIO}
  \item Adaptation of {P}erformer \cite{performer}
  \item Adaptation of {A}udiomer \cite{audiomer} in PyTorch (too time-consuming to port to JAX)
\end{itemize}

Across these experiments, we tried various learning rate schedules,
preprocessing schemes, input embeddings, optimizers (notably RMSProp, Adam and
LAMB), and optimizer hyperparameters. They were carried out primarily using, and
in parallel with the development of, the deep-learning software project
associated with this work. In addition to code specific to the experiments
listed above, it involved developing:

\begin{itemize}
  \item Training, preprocessing \& augmentation utilities, among which:
    \begin{itemize}
      \item A learning rate finder, inspired by \cite{smith2018disciplined}
      \item Multi-GPU training with gradient averaging
      \item Gradient visualization
      \item Stochastic Weight Averaging \cite{swa}
      \item Model checkpointing
      \item Trace preprocessing (point-wise and global)
      \item Training progress monitoring \& associated plots
    \end{itemize}

  \item A reproduction of the Signal-to-Noise ratio analyses shown in
        \cite{ascadv1}, whose code was not included in paper's repository
  \item An exploratory notebook to highlight the properties of Fourier
  positional encoding
  \item Various scripts to train models on the Jean Zay platform
\end{itemize}

We list some of our findings below.

\subsection{Reproducing ASCADv1}

One of the first thing we noticed while trying to reproduce \cite{ascadv1}'s
results was how difficult it was for the network to converge.

We initially thought this was a problem with our code, but even with the
authors' original code, the categorical cross-entropy loss goes from 5.5451 on
training set and 5.5452 on the validation set at epoch 5, to 5.5448 on the
training set and \emph{5.5453} on the validation set at epoch 20.

After prolonging training runs, we found that the original network usually
experienced a sharp drop in the loss around the 30th epoch, in spite of the
learning rate remaining unchanged. However, we did not witness such a drop in
our JAX reimplementation until we added batch normalization \cite{bn} within
convolutional blocks and in the final classifier.  This may be due to minute
differences in the implementations of the RMSProp optimizer between Optax and
Tensorflow, coupled with the network's inherent difficulty to converge on this
task.

\subsection{Reproducing \cite{cnnMethodology}}

As previously mentioned, \cite{cnnMethodology} made no mention of the ASCADv1 variable-key dataset.
We found that, unlike the original ASCADv1 VGG-based network, the authors' best
network for the fixed-key dataset did not converge on the variable-key dataset,
likely owing to its very low complexity, which may be insufficient to model the
interactions between different key bytes and the target variable.

After reimplementing this network within our framework, we experimented with the
learning rate schedule. We quickly noticed that even minute changes in learning
rate at various stages of training would greatly affect the outcome - sometimes
preventing convergence altogether.

\begin{figure}
\begin{subfigure}{.5\textwidth}
  \centering
  \includegraphics[width=0.8\linewidth]{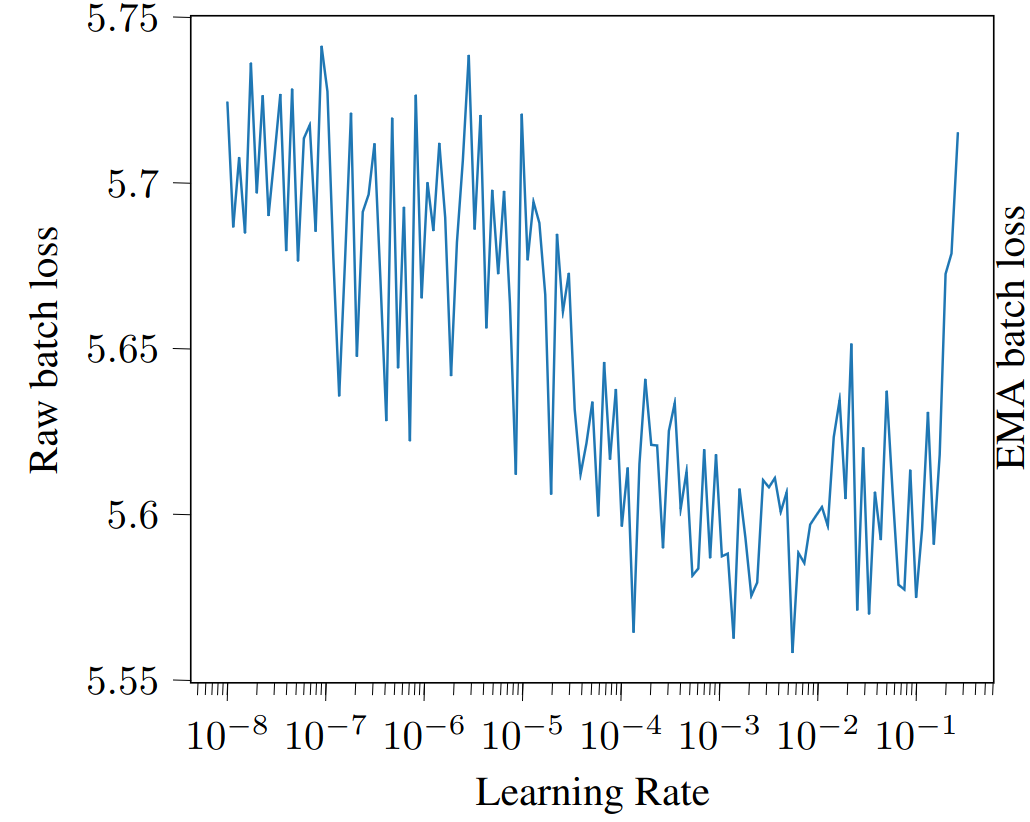}
  \caption{Raw}
\end{subfigure}
\begin{subfigure}{.5\textwidth}
  \centering
  \includegraphics[width=0.8\linewidth]{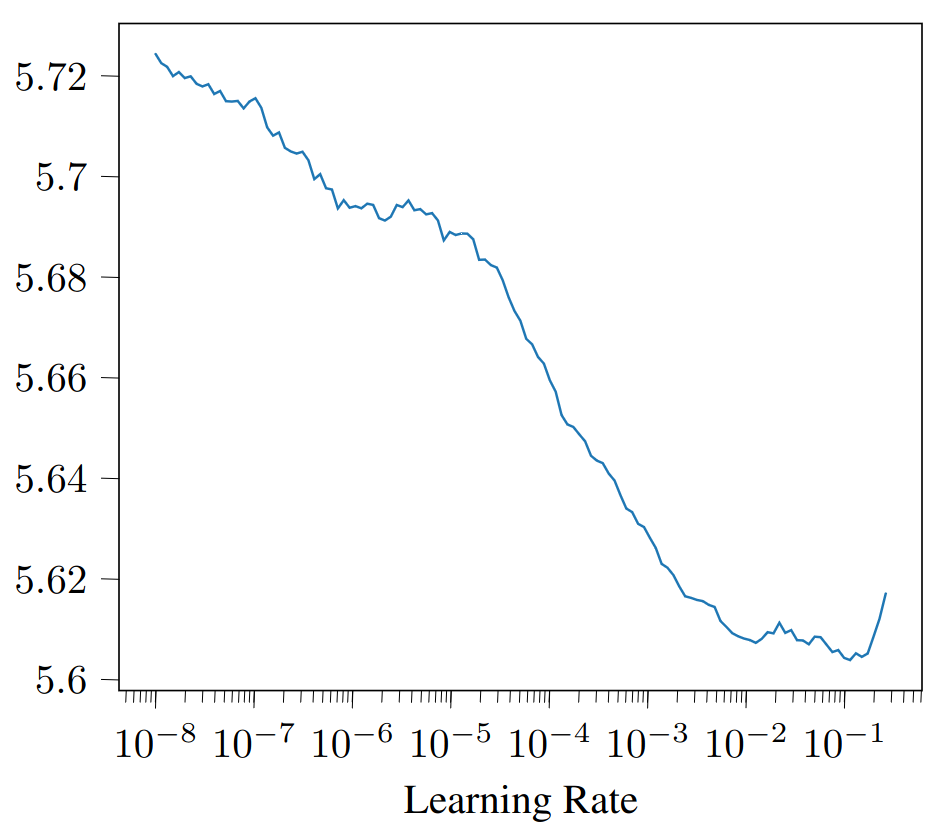}
  \caption{Smoothed}
\end{subfigure}
\caption{Learning rate finder output (EMA = Exponential Moving Average)}
\label{fig:lrfind}
\end{figure}

Since the appropriate learning rate is affected by the dataset and architecture,
we decided to implement a learning rate finder, in order to be able to get LR
upper bounds to use with custom schedules. We could not find a ready-made JAX
implementation, so we wrote our own. It is loosely based on the approach
described in \cite{lrSmith}, implemented in the fastai library \cite{fastai}.
The learning rate is increased exponentially (as in fastai, and unlike in
\cite{lrSmith}, where the increase is linear) over one epoch; the maximum usable
learning rate is then chosen as the maximum value before which the loss
stagnates or increases.  When it starts exploding, the graph is automatically
truncated for readability.  See \autoref{fig:lrfind}. We used this tool
pervasively in our notebooks throughout the rest of our experiments.

One we had a learning rate finder, we started experimenting with various
schedules, trying to achieve super-convergence \cite{superconv}. We tried a
cosine schedule, but noticed that its peaks hurt the network's performance.  We
eventually settled for an exponentially-decayed cosine schedule, with a period
of one-fifth of the total training time, and a half-life of half the total
training time. We observed significantly lower (on the order of 50\% lower)
training and test set losses with this setup when training over 50 epochs,
compared to the one-cycle schedule used in \cite{cnnMethodology}, and
accordingly lower guessing entropy. We were able to do so even at a batch size
of 400 (with which we scaled the learning rate linearly), with across 8 GPUs.

\begin{figure}
  \centering
  \includegraphics[width=0.8\textwidth]{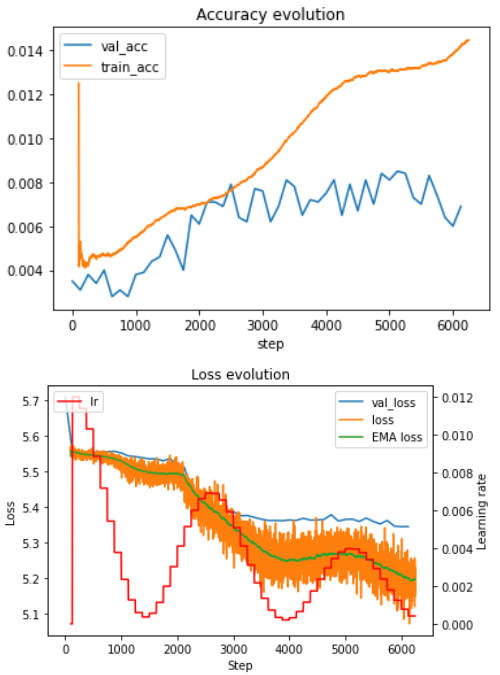
}
\caption{Training plot of our reimplementation of \cite{cnnMethodology}'s best
CNN in the synchronized cast on ASCADv1. The bidirectional impact of the
learning rate on the network's ability to learn can be seen. "EMA loss" is an
exponential moving average of the training loss to account for its high variance
across batches.}
\label{fig:customsched}
\end{figure}

We plot the network's training curve using this schedule on \autoref{fig:customsched}.

\subsection{Is Categorical Cross-Entropy the right loss function?}

While running experiments, we noticed a peculiar phenomenon, which was also
described in \cite{PART}: it is possible for our networks to keep improving with
regards to their guessing entropy, while the validation loss goes \emph{up} and
the training loss goes down - which would normally indicate overfitting.

We hypothesize that this is due to categorical cross-entropy being an inadequate
choice of loss function for the problem at hand. It does not adequately model
the impact of outliers on the final guess, nor of the importance of the
relative order of guesses, moreso than the raw magnitude of their log likelihood.

We found the proposal in \cite{rankingLoss} of a loss function
specifically tailored to the SCA context alluring, but ran into issues when
trying to reproduce results using the project's GitHub repository, as the code could not run as-is.
Due to time constraints, we did not attempt to port it to JAX.

\subsection{Training Transformers}

A list of Transformer architectures we explored was given above. The tests
conducted with them were particularly resource-intensive, always requiring us to
use as many GPUs as possible in parallel in order to get feedback at an
acceptable pace.

\paragraph{Perceiver IO} Adapting Perceiver IO \cite{perceiverIO} was relatively
straightforward, as the project's GitHub repository comes with an encoder suited
for audio processing that we were able to repurpose. As the model is known to be
able to directly attend to individual pixels in the ImageNet setting, we used
patches of size 1.  We used Fourier positional encoding with 64 bands, as well
as learnable encoding of a comparable size. We initially used the project's
default optimizer settings and learning rate scheduling scheme. As the loss did
not decrease to a satisfactory value, even on the training set, we lowered the
dropout rate to 0 and attempted various maximum learning rates, to no avail: we
still observed no convergence with Fourier embeddings. The model did, however,
converge with learnable embeddings, but with poor generalization.

\paragraph{Performer} For Performer \cite{performer}, we extracted the FAVOR+
mechanism (of which an implementation was published by Google) and built our own
Transformer on top of it.  We used two to six Transformer blocks, with four to
eight heads, and a query/key/value size of 64 to 128. For input embedding, we
either extracted patches of the input traces as-is, or used a shallow set of
convolutional layers to encode them.  Once again, Fourier embeddings positional
were ineffective. With learnable positional embedding, the model did learn on the
training set, but did not generalize to the validation set within 50 epochs.

\paragraph{Audiomer} We did a summary test with the Audiomer \cite{audiomer}
architecture. Because of tight coupling between the model's architecture and its input size,
we could not adapt it to our 1400-dimensional traces in a straightforward way.
As a result, we resampled the traces to the length of 8192 timesteps
expected by the model. The model's loss diverged with its default precision,
which used 16-bit floats. With 32-bit floats, it remained stable. 
However, we found that, with other parameters left at default values and a max learning
rate of $10^{-3}$, the model did not learn to be better than a random classifier
after 300 epochs, even on the training set. We realize that resampling may have
introduced artefacts, but the lack of convergence may also indicate that
regularization was too strong (with a dropout rate of 0.2), or that the model's
complexity was too low.

In spite of our efforts, the results we obtained were not encouraging. Given our
previous observations regarding the difficulty of achieving convergence, it
seems that application of this family of models to the problem may be
ineffective without careful architectural considerations, and further
hyperparameter tuning with regards to the length of the optimizer's warm-up
period, the weight decay factor, Adam/LAMB's $\beta_1$ and $\beta_2$ parameters,
the dropout rate, etc. This is especially likely as Transformers are known to be
inherently delicate to train \cite{trtr}.

\section{Conclusion and future work}

We developed our own software project to perform Side-Channel Attacks with deep learning using JAX.
With it, we were able to reproduce architectures proposed in \cite{ascadv1} and
\cite{cnnMethodology}. We optimized aspects of the training process, such as
learning rate scheduling, and conducted our own data exploration (gradient
visualization, signal-to-noise analyses). We discovered that
\cite{cnnMethodology}'s models did not generalize to a methodologically-rigorous
setting in which the key is unknown on the attack dataset. We tried applying
various Transformer-based models on the ASCADv1 variable-key dataset, with limited success.

More careful input preprocessing may be required in order to apply Transformers
to the SCA context.  We encourage the investigation of techniques such as STFTs,
especially for high-resolution and raw traces, as spectral representation is
often advantageous for deep learning in a signal-processing context
\cite{gongAST, tinytrans}.  We also highlight wavelet decomposition
\cite{waveletSCA} as a possible lead to follow.

In parallel, instead of trying to port the problem to a full Transformer
architecture, we suggest trying to incorporate attention gradually into
state-of-the-art SCA models.  We did not have time to pursue our original goal
of building directly upon \cite{PART}; similarly to the approach detailed in it,
we suggest attempting to use a locally-connected layer as a means of embedding
individual clocks cycles before feeding them to a Transformer.

\section*{Code Availability}

Our code was made available to the team we worked with at Telecom Paris's LTCI,
in the form of a private GitHub repository and corresponding environment on the
Jean Zay platform, so that it may be used for further research.

\section*{Acknowledgements}
Thanks to Prof. Laurent Sauvage for his supervision and guidance, and to Arnaud
Varillon for insights into side-channel attacks.

\section*{Update History}
The primary content of this manuscript was completed on February 3, 2022, and \href{https://archive.ph/DBUD3}{originally appeared} on the website of the institution's Embedded Systems Concentration. Any revisions after this date pertain to formatting and minor corrections.

\printbibliography

\end{document}